\documentclass[aps,prl,showpacs,amsmath,amssymb,superscriptaddress,preprint]{revtex4-1}

\usepackage{graphicx}
\usepackage{amsmath}
\usepackage{hyperref}
\usepackage{dcolumn}
\usepackage{bm}
\usepackage{color}
\usepackage{nicefrac}

\begin{document}

\title{Higher cumulants of baryon number in critical QCD}

\author{N.~G.~Antoniou}
 \email[]{nantonio@phys.uoa.gr}
\affiliation{Faculty of Physics, University of Athens, GR-15784 Athens, Greece}

\author{F.~K.~Diakonos}
 \email[]{fdiakono@phys.uoa.gr}
\affiliation{Faculty of Physics, University of Athens, GR-15784 Athens, Greece}

\author{N.~Kalntis}
 \email[]{nikos.kalntis@gmail.com ; sph1300066@uoa.gr}
 % \email[]{sph1300066@uoa.gr}
\affiliation{Faculty of Physics, University of Athens, GR-15784 Athens, Greece}

\author{A.~Kanargias}
\email[]{akanaryias@hotmail.com ; sph1300071@uoa.gr}
%\email[]{sph1300071@uoa.gr}
\affiliation{Faculty of Physics, University of Athens, GR-15784 Athens, Greece}

\date{\today}

\begin{abstract}
\noindent
We study the higher moments of the baryon number in the immediate neighbourhood of the QCD critical endpoint within the framework of Ising-QCD thermodynamics (N.~G. Antoniou {\it et al},
%, F.~K. Diakonos, X.~N. Maintas and C.~E. Tsagkarakis, 
arXiv:1705.09124 [hep-ph]). We show that the kurtosis, as a function of the freeze-out baryon chemical potential, attains a sharp minimum very close to the critical point. We argue that the sharpness of this minimum is due to the narrowness of the critical region in the chemical potential direction. Our analysis reveals that the broad minimum of the kurtosis observed in Au+Au central collisions at STAR (in RHIC-BES I) in the colliding energy region $17$ GeV $< ~\sqrt{s}~<$ $39$ GeV is apparently only a precursor of the critical point and not a signature of its location.

\end{abstract}

%\pacs{} 

\maketitle

One of the basic experimental strategies in the search for the QCD critical endpoint is the determination of higher cumulants of conserved quantities, like baryon number, for a wide range of energies. This program is followed in the Beam-Energy Scan (BES-I) at the STAR experiment (BNL-RHIC) using Au+Au collisions with colliding energies varying in the range $7$ GeV $\leq ~ \sqrt{s} ~\leq$ $200$ GeV \cite{RHIC,Luo2017}. A central quantity considered in these studies is the baryon-number kurtosis times the corresponding variance squared, denoted as "$\kappa \sigma^2$", which is argued to be independent of the fireball's size \cite{Stephanov2009,Athanasiou2010}. The reason for considering this quantity is that there are theoretical considerations claiming that $\kappa \sigma^2$ attains a non-monotonic behaviour, as a function of the energy, within the critical region \cite{Stephanov2009}. This non-monotonic behaviour of $\kappa \sigma^2(\sqrt{s})$ is broadly recognized as a strong signature for the occurrence of critical fluctuations. Furthermore, recent experimental results of RHIC-BES on $\kappa \sigma^2$ measurement show that the function $\kappa \sigma^2 (\sqrt{s})$ displays a broad minimum in the region $17$ GeV $< ~\sqrt{s}~<$ $39$ GeV which is interpreted to be consistent with the presence of a critical point nearby \cite{Luo2017}. 

The aim of the present letter is twofold: firstly we will demonstrate that considering the existing $\kappa \sigma^2$ data from STAR experiment at RHIC-BES I as a function of the freeze-out baryochemical potential $\mu_B$, and not the colliding energy $\sqrt{s}$, we find indications for the occurrence of (at least) two minima instead of one. This observation weakens the argumentation connecting non-monotonic behaviour of $\kappa \sigma^2(\sqrt{s})$ with the emergence of critical fluctuations, since the baryochemical potential is a monotonic function of $\sqrt{s}$ \cite{Andronic2010}. Secondly, employing the Ising-QCD partition function introduced in \cite{Antoniou2017}, we will show that  the appropriately defined non-Gaussian component of kurtosis displays a sharp minimum, with respect to the scale of the energy scanning, as a function of $\mu_B$, close to the critical value $\mu_c$. The sharpness of this minimum is fundamental in nature, induced by the universality class (3d-Ising) of the associated transition. Thus, our treatment shows that the non-monotonic behaviour alone is not a sufficient condition for observing critical fluctuations. According to our findings the broad minimum observed in Au+Au collisions at RHIC can be described by a standard $\phi^4$ effective action showing no traces of a phase transition. It is rather a precursor of the critical point while the critical region itself is characterized by a sharp minimum described by the 3d Ising effective action valid in the immediate neighbourhood of the critical point \cite{Tsypin1994,Antoniou2017}.   

We start our analysis recalling the Ising-QCD effective action introduced in \cite{Antoniou2017} for the description of the baryon number density $n_b(\mathbf{x})=\phi(\mathbf{x}) \beta_c^3$ fluctuations within the critical region:
\begin{eqnarray}
S_{eff}&=&\int_V d^3 \hat{\mathbf{x}} \left[ \frac{1}{2} \vert \hat{\nabla} \phi \vert^2 + U(\phi) - \hat{h} \phi \right]\nonumber \\
U(\phi)&=&\frac{1}{2} \hat{m}^2 \phi^2 + \hat{m} g_4 \phi^4 + g_6 \phi^6
\label{eq:1}
\end{eqnarray}
All quantities occurring in Eq.~(\ref{eq:1}) including the field $\phi$ are dimensionless. In particular $g_4 \approx 1$, $g_6 \approx 2$ are universal constants estimated in \cite{Tsypin1994} while $\hat{m}=\beta_c \xi^{-1}$, ($\xi$ being the correlation length) and $\hat{h}=(\mu_B - \mu_c) \beta_c$ is the ordering field. The action (\ref{eq:1}) can be used to calculate the  Ising-QCD partition function:
\begin{equation}
\mathcal{Z}=\sum_{\{ \phi \}}  \exp(-S_{eff})
\label{eq:2}
\end{equation}
summing over long wavelength modes of $\phi$ (constant configurations, see \cite{Antoniou2017} for details) and subsequently the moments of the baryon number $\langle N^k \rangle$ following the treatment  in \cite{Antoniou2017}. At the critical point $\mu_B = \mu_c$, $T=T_c$ it is found that the first moment $\langle N \rangle$ scales with the size $M=\frac{V}{V_0}$ ($V_0$ being the approximate volume of a single baryon) as:
\begin{equation}
\langle N \rangle \sim M^{q}~~~~;~~~~~q=\frac{5}{6}~(\mathrm{3d-Ising})
\label{eq:3}
\end{equation}
Furthermore, defining as critical region the domain in the $(\ln \zeta = \frac{\mu_B - \mu_c}{k_B T_c}, t= \frac{T-T_c}{T_c})$ plane for which the scaling relation:
\begin{equation}
\langle N \rangle \sim M^{\tilde{q}}~~~;~~~3/4 < \tilde{q} < 1
\label{eq:4}
\end{equation}
holds, it was found that the critical region is very narrow along the baryochemical potential direction ($\Delta \mu_B \approx 5$ MeV for $\hat{m}=0$). A result which is consistent with experimental findings from intermittency analysis of proton transverse momenta at SPS (NA49 experiment). This small value of $\Delta \mu_B$ is related to the  
fact that the exponent in the scaling relation (\ref{eq:3}) is very close to one. To demonstrate this, we adopt a more general form for the effective action (\ref{eq:1}) at the critical temperature ($\hat{m}=0$):
\begin{equation}
\tilde{S}_{eff}=\int_V d^3 \hat{\mathbf{x}} \left[ \frac{1}{2} \vert \hat{\nabla} \phi \vert^2 + g \phi^{\delta +1} - h \phi \right]
\label{eq:5}
\end{equation}
with $g$ an arbitrary constant and $\delta$ the isothermal critical exponent. With this choice we recover the 3d-Ising case for $\delta=5$, $g \approx 2$. Using the effective action (\ref{eq:5}) we calculate the first moment $\langle N \rangle$ as a function of the size $M$ for various values of $\delta$, in the range $1 \leq \delta \leq 5$, and $\ln \zeta$. Since we are interested in the scaling behaviour $\langle N \rangle \sim M^{q_{\delta}}$ the exact value of $g$ is irrelevant. Thus, we assume $g \approx 2$ without loss of generality. In Fig.~1a we show the result of this calculation. We plot the scaling exponent $q_{\delta}$ as a function of $\ln \zeta$. The various curves correspond to different choices of the exponent $\delta$. In the plot each curve is labelled by the value of $q_{\delta}$ at $\mu_B=\mu_c$ ($\ln \zeta =0$) which we denote $q_{\delta,0}$. We observe that the curve corresponding to $q_{\delta,0}={5 \over 6}$ (3d-Ising universality class) is very abrupt and decreasing $q_{\delta,0}$ (decreasing $\delta$) the associated curves become more and more smooth. Therefore for a given interval of $q_{\delta}$-variation, the corresponding range of chemical potential values around the critical one is very narrow for $q_{\delta,0}={5 \over 6}$, and broadens with decreasing $q_{\delta,0}$. This is clearly displayed in Fig.~1b where we present the chemical potential range $\Delta \mu_B$ needed to cover the region ${3 \over 4} < q_{\delta} < 1$ as function of $q_{\delta,0}$. 

\begin{figure}[tbp]
\centering
\includegraphics[width=0.55\textwidth]{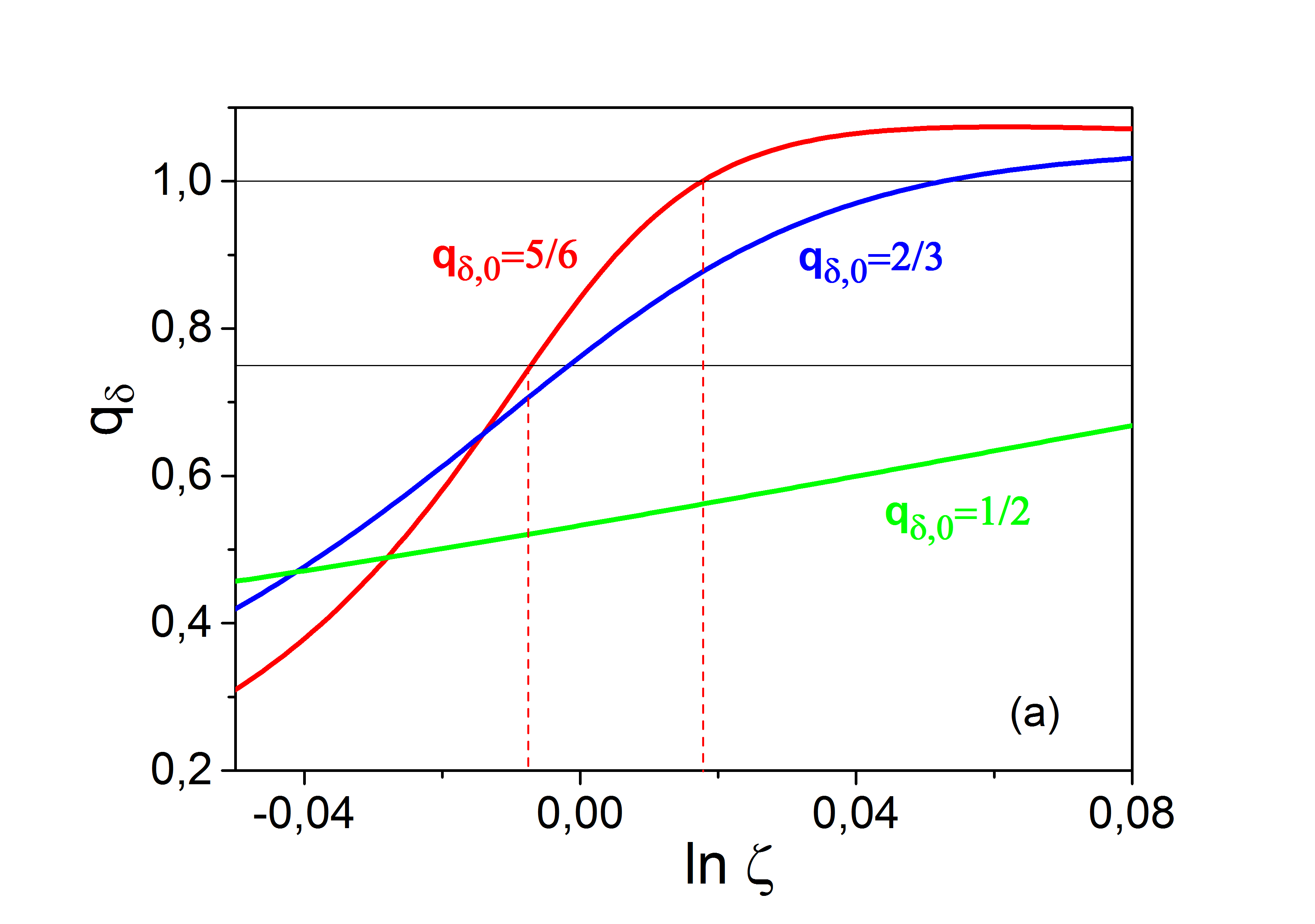}
\includegraphics[width=0.55\textwidth]{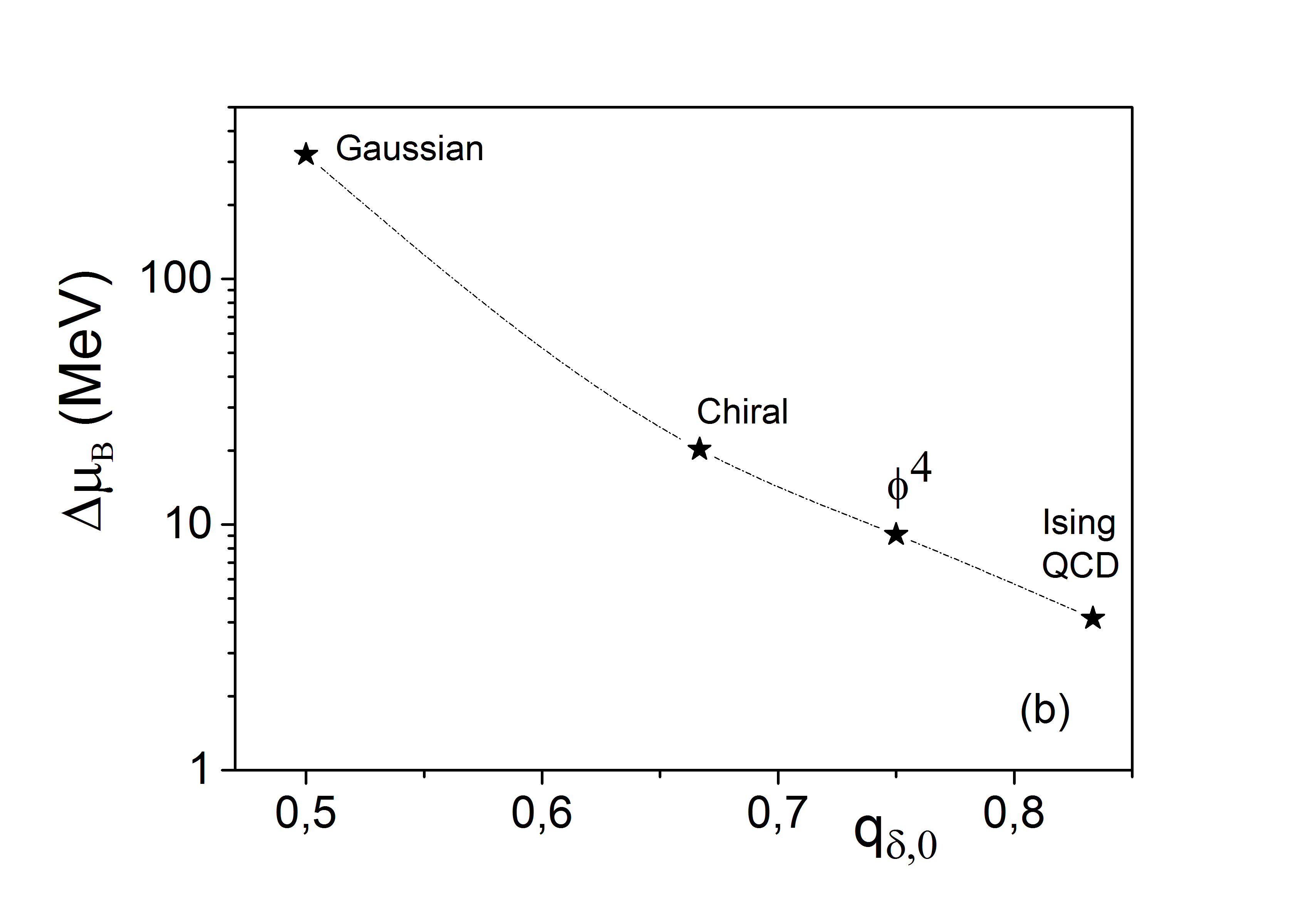}
\caption{(a) The exponent $q_{\delta}$ as a function of $\ln \zeta = {\mu_B - \mu_c \over k_B T_c}$ for various values of $\delta$. The curves are labelled by $q_{\delta,0}$ being the value of $q_{\delta}$ at $\ln \zeta = 0$. To illustrate the narrowness of the critical region we use the horizontal black solid lines and the vertical red dashed lines which determine the region where ${3 \over 4} \leq q_{\delta} \leq 1$ for the 3d-Ising case ($\delta=5$). In (b) we plot the chemical potential interval $\Delta \mu_B$ for which ${3 \over 4} < q_{\delta} < 1$ as a function of $q_{\delta,0}$ assuming $T_c = 167$ MeV. The dotted line is used to guide the eye.} 
\label{fig:fig1}
\end{figure} 
 
Thus, according to Fig.~(\ref{fig:fig1}) the narrowness of the critical region along the chemical potential direction is of fundamental origin, being dictated by the universality class of the transition via the isothermal critical exponent $\delta$ and not by the details of the description.

The next point in our analysis is the calculation of the non-Gaussian kurtosis in the critical region employing the partition function (\ref{eq:2}). We use the definition:
\begin{equation} 
\kappa=\frac{C_4}{C_2^2}-3~~~;~~~C_2 =\langle (\delta N)^2 \rangle,~C_4 =\langle (\delta N)^4 \rangle
\label{eq:6}
\end{equation}
with $\delta N = N - \langle N \rangle$ ($N$=baryon number) and $C_2$, $C_4$ the second and fourth order cumulants respectively. As usual in Eq.~(\ref{eq:6}) the Gaussian contribution is subtracted. The calculation is done for $T=T_c$ ($\hat{m}=0$) while $\ln \zeta$ is varied in the region $-0.05 < \ln \zeta < 0.05$. As shown in Fig.~1a the critical region, estimated using the finite-size scaling behaviour of $\langle N \rangle$, is $ -0.01 < \ln \zeta < 0.02$ (it is the region determined by the red dashed lines), leading to a chemical potential range $\Delta \mu_B \approx 4.5$ MeV displayed in Fig.~1b (see also \cite{Antoniou2017}). The result of the kurtosis calculation is plotted in Fig.~2 showing a clear minimum at $\mu_B \approx \mu_c$. The red arrows indicate the critical region. 
 
\begin{figure}[tbp]
\centering
\includegraphics[width=0.55\textwidth]{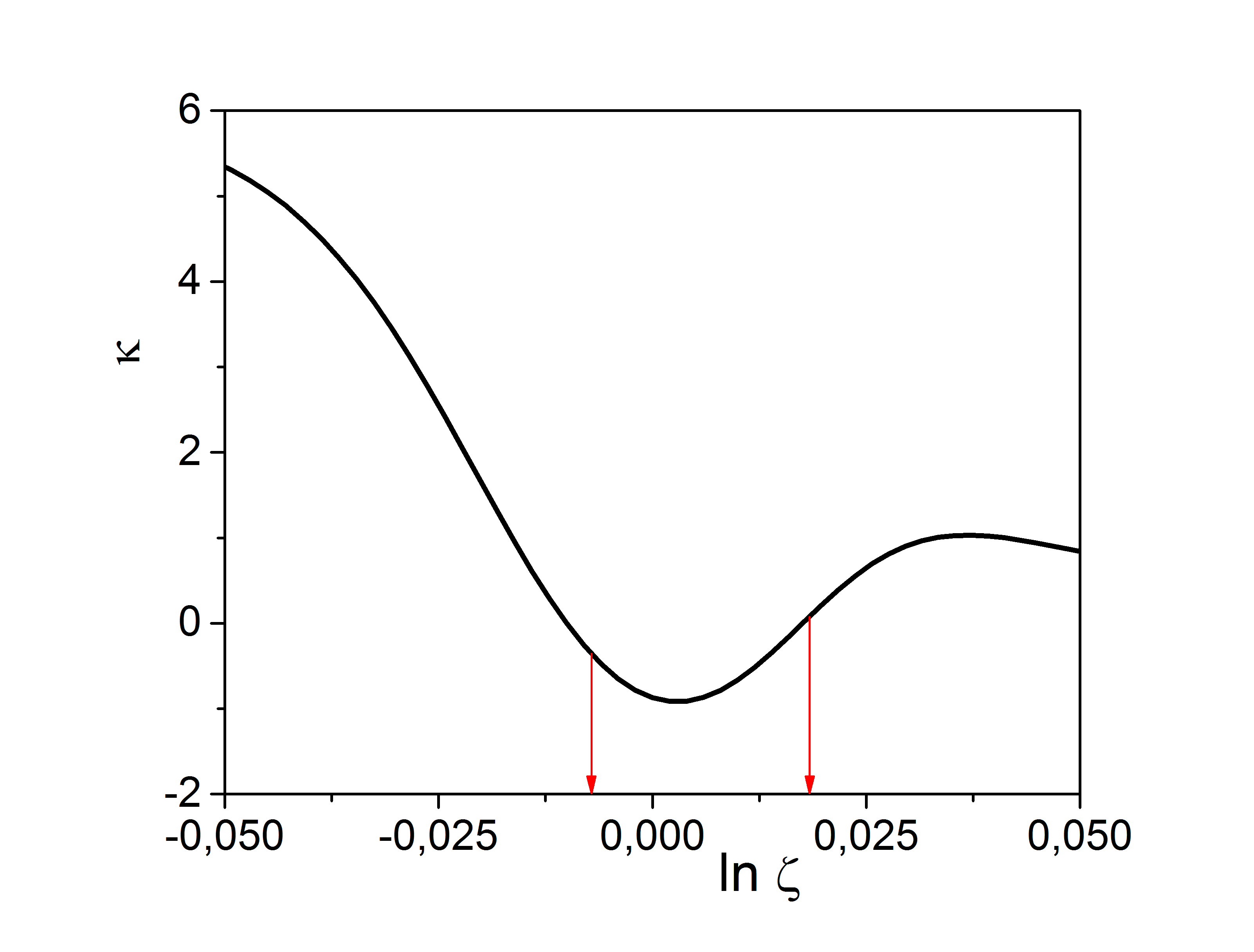}
\caption{The Ising-QCD prediction for the non-Gaussian part of the kurtosis of the baryon number as a function of $\ln \zeta = \frac{\mu_B - \mu_c}{k_B T_c}$. The red arrows show the borders of the critical region.} 
\label{fig:fig2}
\end{figure} 

In Fig.~3 we plot the published RHIC-BES I experimental results \cite{Luo2017} for $\kappa \sigma^2$ of the proton-number distribution using as abscissas, instead of the energy $\sqrt{s}$, the corresponding freeze-out chemical potential values. We observe that the function
$\kappa \sigma^2(\mu_B)$ indicates the existence of two minima in the range of chemical potential values scanned by the freeze-out states generated at STAR experiment in RHIC-BES I. It is clear that the same structure is also contained in the plot $\kappa \sigma^2(\sqrt{s})$ since the relation between freeze-out chemical potential and colliding energy is monotonic. However, due to the change of the scale and the existing errors this information is washed out in the plot of $\kappa \sigma^2(\sqrt{s})$. This observation has far reaching consequences. First, it proves that the use of extrema of higher order cumulants (or their ratios) as a unique signature for the critical point is insufficient and may lead to erroneous conclusions, at least concerning the location of the critical point. Second, it indicates that the correct energy scale for exploring the QCD phase diagram in the search for the critical region is the baryochemical potential $\mu_B$ and not the energy $\sqrt{s}$. In fact one can make a step ahead trying to describe the behaviour of $\kappa \sigma^2$ shown in Fig.~3 utilizing a non-critical effective action of the form:
\begin{eqnarray}
\hat{S}&=&\int_V d^3 \hat{\mathbf{x}} \left[ \frac{1}{2} \vert \hat{\nabla} \phi \vert^2 + U[\phi]-\frac{\mu}{k_B T} \phi \right]
\nonumber \\
U[\phi]&=&a_2(\mu_B,T) \phi^2 + a_3(\mu_B,T) \phi^3 + a_4(\mu_B,T)\phi^4 
\label{eq:8}
\end{eqnarray}
where $\phi=n_B \beta_c^3$ and $a_i$, $i=2,3,4$ are dimensionless constants which are suitably tuned to reproduce the experimental values for $\kappa \sigma^2$ in Fig.~3. As in the cases discussed previously the partition function determining the thermodynamics of the baryon number density is given by $\hat{\mathcal{Z}}=\sum_{\{ \phi \}}  \exp(-\hat{S})$ and the summation is over constant $\phi$-configurations. Indeed, this simple model is able to reproduce the experimentally observed behaviour of $\kappa \sigma^2$ as a function of the freeze-out $\mu_B$ and the result is shown by the green and red solid lines in Fig.~3. To describe the data in the region of lower chemical potential values (crossover) the term proportional to $\phi^3$ can be set to zero while for the higher chemical potential values (first order transitions region) this term is large and negative.

In the plot we see that only in the region around $\sqrt{s}=14.5$ GeV this simple description fails. We claim that this is due to the fact that the critical point is close to the freeze-out state of the Au+Au collisions at $\sqrt{s}=14.5$ GeV ($\mu_B \approx 260$ MeV). Thus, although in a distance from the critical point the quantity $\kappa \sigma^2$, being size independent, is an adequate fluctuation measure for the freeze-out states \cite{Luo2017,Huang1987}, when entering the finite-size scaling region the  correlation length of the finite system and its size become entangled. In fact, near the critical point, where long-range correlations prevail, the kurtosis itself, as given in Eq.~(\ref{eq:6}), is size independent, since $C_2 \sim V^{2 q}$, $C_4 \sim V^{4 q}$ \cite{Antoniou2017}. Therefore, in this region, $\kappa \sigma^2$ depends strongly on system's size and it is inadequate for measuring fluctuations. One has to replace $\kappa \sigma^2$ by the kurtosis $\kappa$ which is size-independent within the finite-size scaling region.  
 
This line of arguments is further supported by details of the calculation of $\kappa$, employing the Ising-QCD partition function and shown in Fig.~2. Using the value $\mu_c=256$ MeV for the critical chemical potential value (as estimated in \cite{Antoniou2017} through finite-size scaling analysis) and $T_c=167$ MeV, which is compatible with the most recent Lattice QCD results considering Polyakov loop or strange quark susceptibilities \cite{Gavai2016,Datta2017}, we calculate $\kappa$ for energies around $\sqrt{s}=14.5$ GeV and map the plot shown in Fig.~2 to that shown in Fig.~3 by the blue solid line. We observe that in the critical region $\kappa$ attains a sharp minimum which is fully compatible with the narrowness of the critical region along the chemical potential direction. Thus, the broad minimum of the data for lower chemical potential values can only be considered as a precursor of the critical point and it has no direct relation to it. 

\begin{figure}[tbp]
\centering
\includegraphics[width=0.55\textwidth]{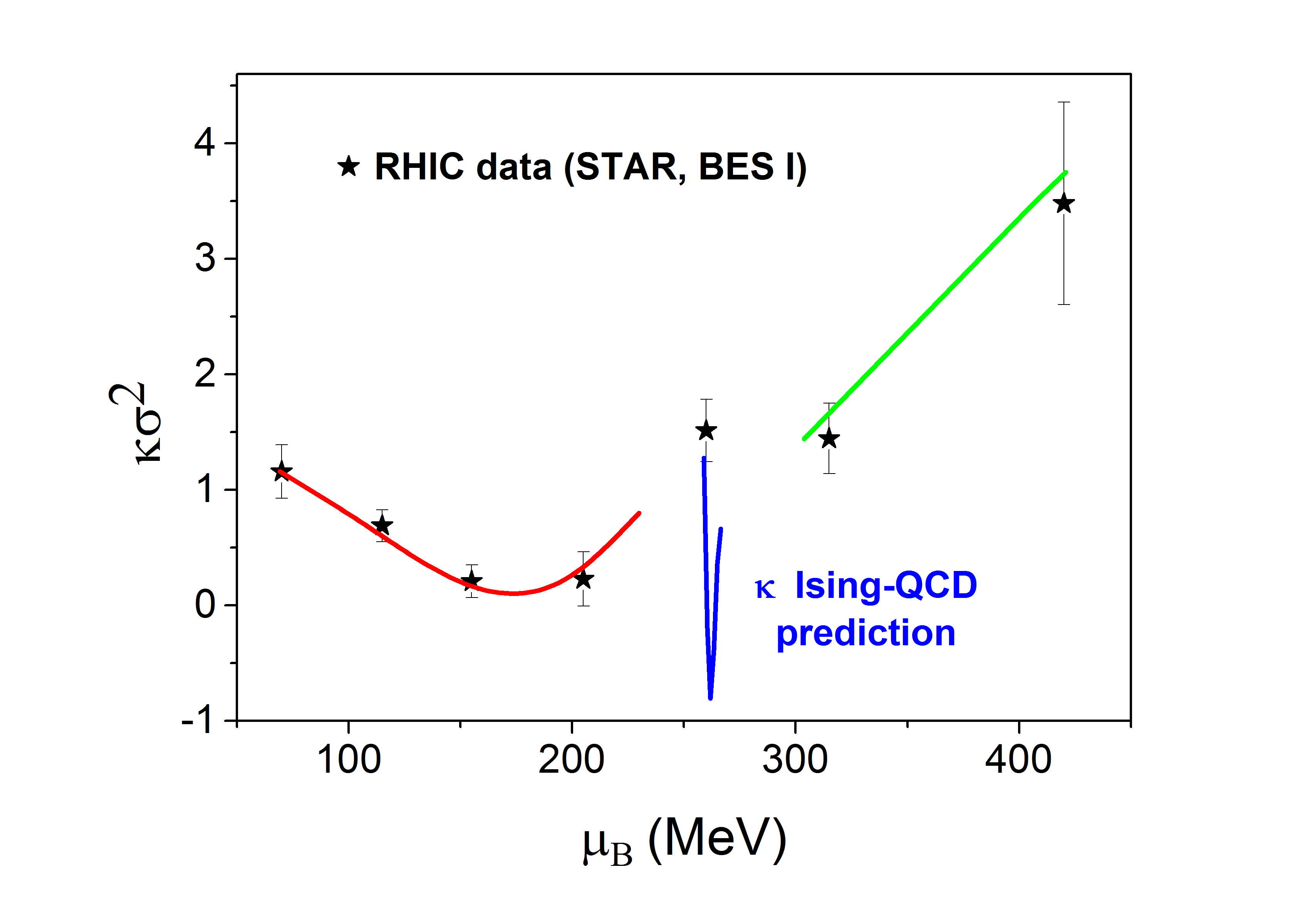}
\caption{Net-proton $\kappa \sigma^2$ as a function of the baryochemical potential $\mu_B$. The stars are the measurements at RHIC-BES I \cite{Luo2017}. The green and red lines are the numerical results using the action (\ref{eq:8}). The blue line is the prediction for $\kappa$ obtained through the Ising-QCD partition function \cite{Antoniou2017}.} 
\label{fig:fig3}
\end{figure} 

A final comment is here in order. In Fig.~4 we plot the Landau free energy $F[\phi]=U(\phi)-\frac{\mu}{k_B T} \phi$ in Eq.~(\ref{eq:8}) for two cases: the red line is representative for the form of the effective potential in the red region (crossover regime) while the green line is representative for the corresponding form in the green region (first order transitions). It is remarkable that this model is consistent with the basic thermodynamic characteristics in the two regions. In the crossover regime there is a single minimum with no additional metastable state while in the first-order regime two non-degenerate shallow minima occur, consistent with the presence of a metastable state. 

\begin{figure}[tbp]
\centering
\includegraphics[width=0.55\textwidth]{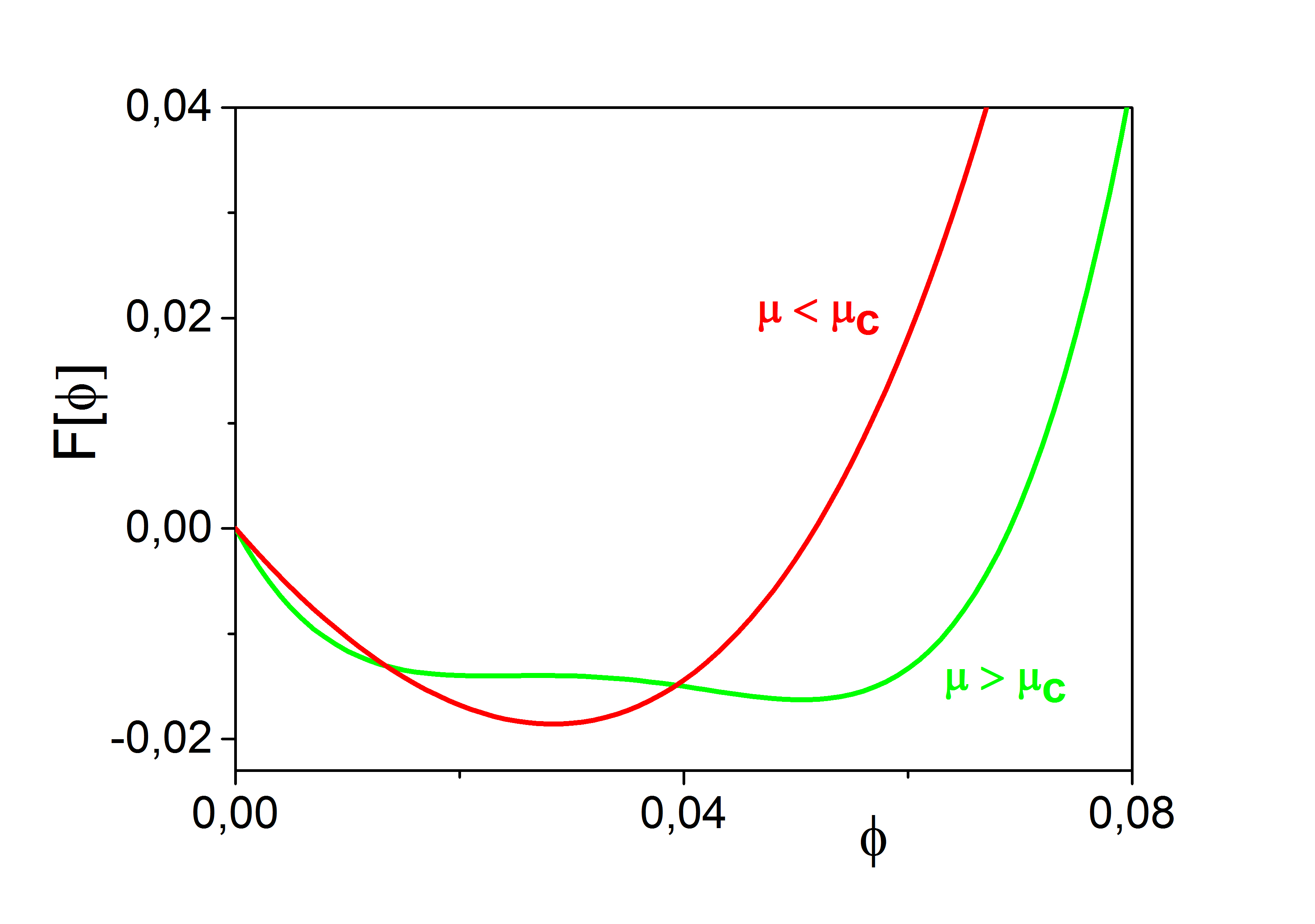}
\caption{The Landau free energy $F[\phi]=U(\phi)-\frac{\mu}{k_B T} \phi$ in (\ref{eq:8}) for $\mu_B < \mu_c$ (red line, crossover regime) and $\mu_B > \mu_c$ (green line, first order transitions region).} 
\label{fig:fig4}
\end{figure} 

It is interesting to determine the (normalized) baryon number distribution in each case as well. The result of this calculation is presented in Fig.~5. As a typical example, for the case $\mu < \mu_c$, we use freeze-out parameters $(\mu,T)$ corresponding to the state generated by Au+Au central collisions at colliding energy $\sqrt{s}=19.6$ GeV. We observe that the shape of the obtained baryon number multiplicity distribution (blue cricles) is fully compatible with the experimental findings in STAR \cite{Luo2017}. The crosses are the result for the baryon number multiplicity distribution in the case $\mu > \mu_c$ using thermodynamic parameters $(\mu,T)$ which correspond to those of the freeze-out state formed by central Au+Au  collisions at $\sqrt{s}=11.5~ GeV$. The mean net-proton number is $\langle N \rangle \approx 30$ in accordance with the published STAR results \cite{Luo2017}. The somewhat distorted baryon-number distribution for $\mu > \mu_c$ reflects the presence of the two non-degenerate shallow minima of $F[\phi]$ in Fig.~(\ref{fig:fig4}). In fact, it is a sign for the appearance of a metastable state in the region of first-order phase transitions. Since the formation of such metastable states is a rather model-independent property  of the thermodynamic potential in this region \cite{Parisi1988,Jiang2017}, one may conjecture that the effect of a distorted multiplicity distribution of net-protons can be detected in studies of baryon-rich matter, with the new, forthcoming facilities FAIR and NICA. 

\begin{figure}[tbp]
\centering
\includegraphics[width=0.55\textwidth]{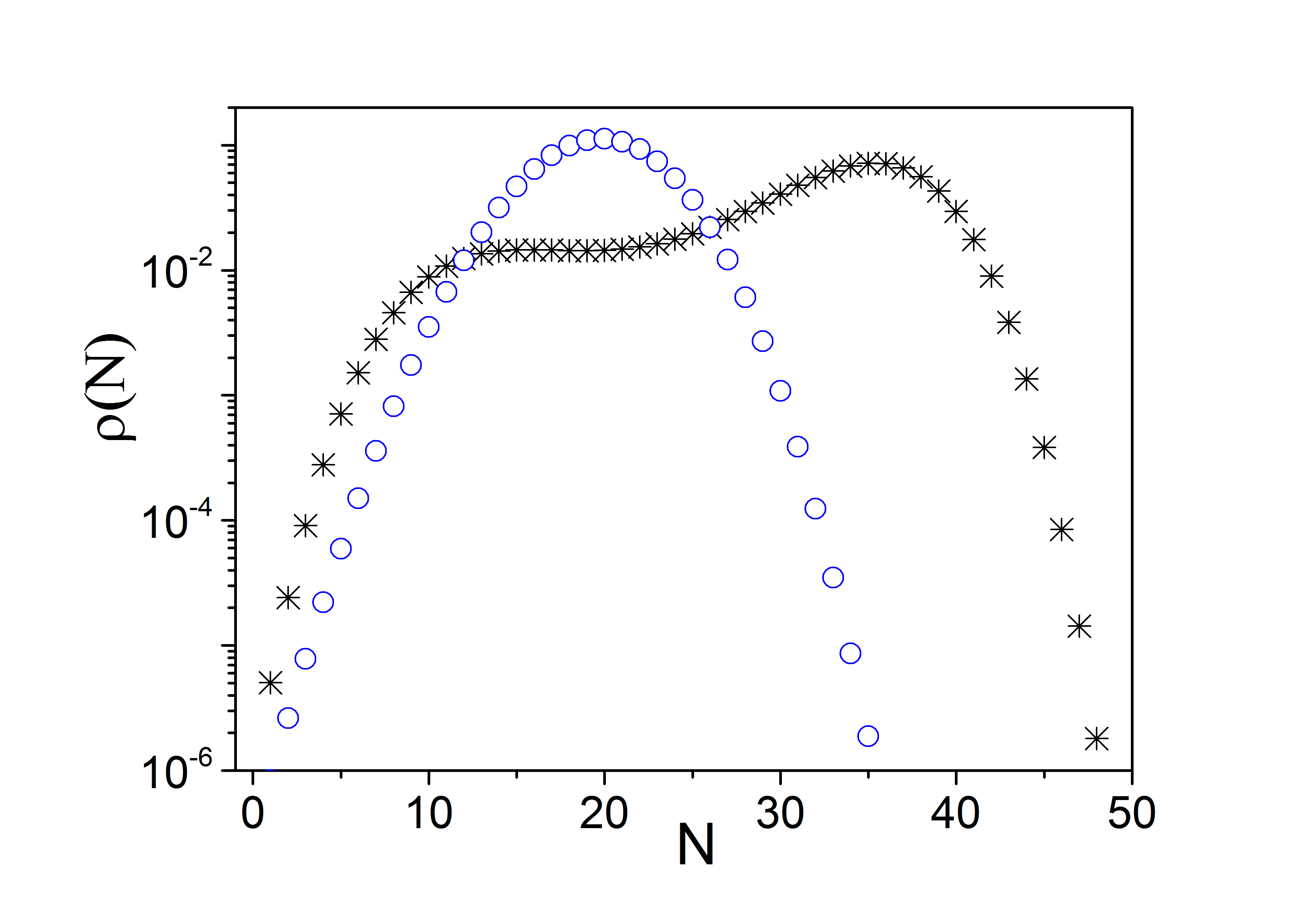}
\caption{The normalized baryon number probability distribution                                 for $\mu_B < \mu_c$ (blue circles) and $\mu_B > \mu_c$ (crosses) calculated employing the action (\ref{eq:8}).} 
\label{fig:fig5}
\end{figure}

In conclusion, we have demonstrated that the recently published 
experimental results for $\kappa \sigma^2$ of net-protons in central Au+Au collisions (RHIC-BES I) when presented as a function of the corresponding freeze-out chemical potential show a behaviour which is  compatible with the presence of two minima lying on each side of the value $\mu_B \approx 250$ MeV corresponding to $\sqrt{s}=14.5$ GeV. The minimum at high energies cannot be directly related to QCD critical point. In fact, we have illustrated how a simple $\phi^4$-thermodynamic potential with smooth coefficients $a_i(\mu,T)$ ($i=2,3,4$) of maximal variation $\frac{\Delta a_i}{\langle a_i \rangle} \leq 1.3$ and without a trace of critical behaviour, may reproduce
the broad minimum of $\kappa \sigma^2$ in the crossover region ($\mu_B \ll \mu_c$) and its large increase at low energies ($\mu_B \gg \mu_c$) in a region of first-order transitions (Fig.~(\ref{fig:fig3})).

Furthermore, using as energy scale the chemical potential values instead of the energy $\sqrt{s}$ allows to resolve the critical region and predict, based on Ising-QCD partition function,  a sharp minimum of $\kappa$ close to $\sqrt{s}=14.5$ GeV. This minimum is identified with the location of the critical point which however cannot be captured by the colliding energy scanning program in its present realization. The sharpness of this minimum is a further manifestation of the narrowness of the critical region in the baryochemical potential direction, as argued also in \cite{Antoniou2017}. We also have provided numerical evidence that the narrowness of the critical region along the chemical potential axis is of fundamental origin related to the universality class (3d-Ising) of the transition and not on the details of the description of critical fluctuations.


\begin{thebibliography}{99}

\bibitem{Antoniou2017} N.~G. Antoniou, F.~K. Diakonos, X.~N. Maintas and C.~E. Tsagkarakis, arXiv:1705.09124 [hep-ph].

\bibitem{RHIC}  M.~M. Aggarwal {\it et  al.} (STAR  Collaboration), arXiv:1007.2613;  STAR Note 0598:  BES-II whitepaper: http://drupal.star.bnl.gov/STAR/starnotes/public/sn0598.

\bibitem{Luo2017} X. Luo and N. Xu, Nucl. Sci. Tech. {\bf 28}, 112 (2017); arXiv:1701.02105 [nucl-ex].

\bibitem{Stephanov2009} M.~A. Stephanov, Phys. Rev. Lett. {\bf 102}, 032301 (2009); M.~A. Stephanov, Phys. Rev. Lett. {\bf 107}, 052301 (2011).

\bibitem{Athanasiou2010} C. Athanasiou, K. Rajagopal and M.~A. Stephanov, Phys. Rev. D {\bf 82}, 074008 (2010).

\bibitem{Andronic2010} A. Andronic, P. Braun-Munzinger and J. Stachel, Nucl. Phys. A {\bf 834}, 237C (2010). 

\bibitem{Tsypin1994} M.~M. Tsypin, Phys. Rev. Lett. {\bf 73}, 2015 (1994).

\bibitem{Huang1987} K. Huang, {\it{"Statistical Mechanics"}}, Wiley, New York (1987).

\bibitem{Gavai2016} R.~V. Gavai, {\it{"The QCD critical point: an exciting Odyssey in the Femto-world"}}, Contemporary Physics {\bf 57}, 350 (2016). 

\bibitem{Datta2017} S. Datta, R.~V. Gavai and S. Gupta, Phys. Rev. D {\bf 95}, 054512 (2017).

\bibitem{Parisi1988} G. Parisi, {\it{"Statistical Field Theory"}}, Addison-Wesley, USA, 1988.

\bibitem{Jiang2017} L. Jiang, J-H. Zheng and H. St\"{o}cker, arXiv: 1711.05339 (nucl-th).


\end{thebibliography}
\end{document}